\documentclass[aps,prl,preprint,superscriptaddress]{revtex4}

\usepackage{amsmath}
\usepackage{amsfonts}
\usepackage{amssymb}
\usepackage{graphicx}

\begin{document}


\title{Double scattering of light from biophotonic nanostructures with short-range order}


\affiliation{Department of Applied Physics}
\affiliation{Department of Ecology and Evolutionary Biology and\\Peabody National History Museum}
\affiliation{Departments of Mechanical Engineering, \\ Chemical Engineering and Cell Biology}
\affiliation{Department of Physics}
\affiliation{Center for Research on Interface Structure and Phenomena\\Yale University, New Haven, CT 06511}

\author{H. Noh}
\affiliation{Department of Applied Physics}
\affiliation{Center for Research on Interface Structure and Phenomena\\Yale University, New Haven, CT 06511}
\author{S. F. Liew}
\affiliation{Department of Applied Physics}
\affiliation{Center for Research on Interface Structure and Phenomena\\Yale University, New Haven, CT 06511}
\author{V. Saranathan}
\affiliation{Department of Ecology and Evolutionary Biology and\\Peabody National History Museum}
\affiliation{Center for Research on Interface Structure and Phenomena\\Yale University, New Haven, CT 06511}
\author{S. G. J. Mochrie}
\affiliation{Department of Applied Physics}
\affiliation{Department of Physics}
\affiliation{Center for Research on Interface Structure and Phenomena\\Yale University, New Haven, CT 06511}
\author{R. O. Prum}
\affiliation{Department of Ecology and Evolutionary Biology and\\Peabody National History Museum}
\affiliation{Center for Research on Interface Structure and Phenomena\\Yale University, New Haven, CT 06511}
\author{E. R. Dufresne}
\affiliation{Departments of Mechanical Engineering, \\ Chemical Engineering and Cell Biology}
\affiliation{Department of Physics}
\affiliation{Center for Research on Interface Structure and Phenomena\\Yale University, New Haven, CT 06511}
\author{H. Cao}
\affiliation{Department of Applied Physics}
\affiliation{Department of Physics}
\affiliation{Center for Research on Interface Structure and Phenomena\\Yale University, New Haven, CT 06511}

%



\begin{abstract}

We investigate the physical mechanism for color production by isotropic nanostructures with short-range order in bird feather barbs. While the primary peak in optical scattering spectra results from constructive interference of singly-scattered light, many species exhibit secondary peaks with distinct characteristic. Our experimental and numerical studies show that these secondary peaks result from double scattering of light by the correlated structures. Without an analog in periodic or random structures, such a phenomenon is unique for short-range ordered structures, and has been widely used by nature for non-iridescent structural coloration.


\end{abstract}


\maketitle


Structural colors have attracted much attention in a wide variety of disciplines \cite{Kinoshita2008}. They originate from physical interactions of light with nanostructures. Recent studies have focused on ordered structures in the natural world which produce iridescent colors \cite{Ghiradella1998,Vukusic2000,Parker2001}. However, nature has also used extensively quasi-ordered structures to create vivid colors that are weakly iridescent, e.g., the spongy nanostructures in the feather barbs of numerous birds \cite{Prum2006}. In the 19th century, lack of pronounced iridescence led to the hypothesis that the colors were produced by wavelength selective scattering from single particles. Raman was the first to cast doubt on this hypothesis  \cite{Raman1934}, which was later falsified by Dyck \cite{Dyck1971}. Prum et al. proved the existence of structural order by Fourier analysis of electron micrographs of medullary keratin in feather barbs \cite{Prum1998}. In the last ten years, the previously unappreciated class of quasi-ordered nanostructures have been identified also in bird skins, mammal skins, dragonfly cuticles, and butterfly scales \cite{Kinoshita2008,Prum2006}.

Although it is now recognized that light scattering is organized to contribute effectively to coloration of quasi-ordered structures, there are significant spectral features whose origin remains elusive \cite{Osorio2002,Prum2006,Shawkey2009,Dufresne2009}. In previous studies, optical reflectance spectra are predicted from the structural Fourier spectra with the assumption of single scattering \cite{Shawkey2009}. The dominant length scale of structural correlation gives one peak in the predicted reflectance spectrum. However, many species we have measured feature two reflection peaks, in agreement with prior observations \cite{Osorio2002,Prum2006,Barreira2008}. What is the origin of the second peak? Does multiple scattering play a significant role in coloration? How does the coloration of quasi-ordered structures differ from that of ordered structures? Answering these questions is essential to identify the physical mechanism for coloration of the quasi-ordered structures that are widely distributed in nature and clearly function in animal communication. Moreover, understanding the effects of short-range structural order on multiple scattering of light is fundamental to the physics of light interaction with complex media. In practical terms, the wavelength dependence of light scattering from quasi-ordered structures has not been taken advantage of technologically, and may have useful advantages for photonic coatings across many areas such as the production of color for paint, cosmetics and textile industries and photon management in solar cells.



In this Letter, we investigate the physical mechanism for coloration of nanostructures with short-range order in bird feather barbs by performing angle- and polarization-resolved scattering spectrometry. The primary peak in optical scattering spectra maintains the polarization of incident light and its spectral shift with observation angle agrees with single scattering predictions. The short-range structural order leads to phase correlation of scattered waves and their constructive interference produces a narrow spectral peak. For many avian species, the scattering spectra have a secondary peak which is depolarized and has different angular dispersion from the primary peak. We show that the secondary peak is caused by interference of light scattered twice by the nanostructures. This result is unexpected from the general belief that multiple scattering of light simply broadens the single scattering peak. There is no analogous phenomenon in periodic structures where multiple diffraction or high-order diffraction does not create additional peaks.

\begin{figure}[htbp]
\center
\includegraphics[width=3.4in]{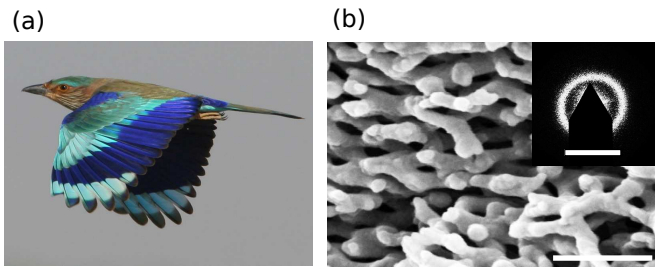}
\caption[SEM]{(a) {\it Coracias benghalensis} (Indian Roller) whose wing feather color was studied by Raman in the 1930s. Photo credit to Pascal Rapin (reproduced with permission) (b) Scanning electron micrograph of the feather barbs showing channel-type nanostructures with $\beta$-keratin rods and air channels. The scale bar = 500nm.  Inset: small-angle X-ray scattering (SAXS) pattern of the feather barbs that exhibits a ring. The scale bar = 0.05 nm$^{-1}$ }
\label{fig:SEM}
\end{figure}

The color-producing quasi-ordered nanostructures within the medullary cells of bird feather barb rami belong to two morphological classes\cite{Dyck1976,Prum2006}. Channel-type nanostructures consist of $\beta$-keratin bars and air channels in elongate and tortuous forms. Sphere-type nanostructures consist of spherical air cavities in a $\beta$-keratin matrix. Our recent studies suggest these nanostructures are self-assembled during phase separation of $\beta$-keratin protein from the cytoplasm of the cell \cite{Dufresne2009}.
In this work we have performed small angle X-ray scattering (SAXS) and optical measurements on both types of structures from many bird species. Since the results are similar, we present here the data from turquoise wing feathers of {\it Coracias benghalensis} ({\it C.  benghalensis}), the same bird species that Raman studied over seventy years ago [Fig. \ref{fig:SEM}(a)].

\begin{figure}[htbp]
\center
\includegraphics[width=3.4in]{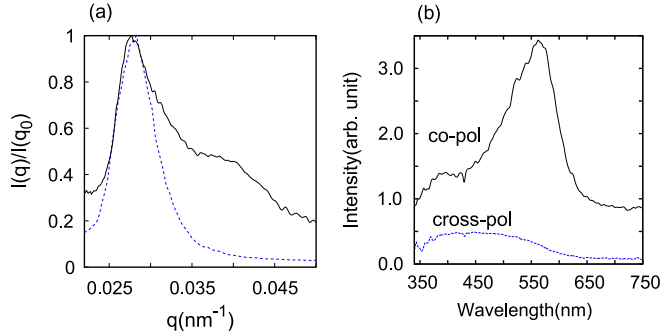}
\caption[polarization]{(a) Azimuthal average of SAXS intensity (blue dashed line) in comparison to nearly backscattered light intensity at normal incidence (black solid line) plotted as a function of the magnitude of spatial vector $q$. The optical scattering spectrum has a secondary peak.
(b) Scattered light intensity of co-polarization (black solid line) and cross-polarization (blue dashed line) as a function of wavelength. The linearly polarized white light is incident at normal, $\phi = 0^{\circ}$. The detector collects the scattered light at an angle of $\theta = 10^{\circ}$ from the surface normal. The polarization direction of the incident light is parallel to the scattering plane that is formed by the surface normal and detector. Similar results are obtained when the polarization of incident light is perpendicular to the scattering plane. }
\label{fig:pol}
\end{figure}

The barb nanostructures of {\it C.  benghalensis} are a 3D network of keratin rods of $\sim$100 nm diameter and air channels of similar width [Fig. \ref{fig:SEM}(b)]. The SAXS pattern exhibits a ring [inset of Fig. \ref{fig:SEM}(b)], implying the nanostructure is isotropic.  Exploiting the rotational symmetry, we average the scattering intensity azimuthally to obtain the intensity $I$ as a function of the magnitude of spatial vector $q$ [Fig. \ref{fig:pol}(a)]. The strong peak of $I(q)$ at $q_0$ = 0.028 nm$^{-1}$ reveals the existence of a dominant length scale $s$ for structural correlation. $s= 2 \pi / q_0$ = 225 nm is approximately half of the wavelength of blue light.
The full width at half maximum (FWHM) $\Delta q$ of the peak in $I(q)$ reflects the range $\xi$ of spatial order, $\xi = 2 \pi / \Delta q$ = 1.16 $\mu$m. Since $\xi$ is only 5 times of the spatial period $s$, the order is short-ranged. Such structures can be called amorphous photonic materials, in analog to amorphous electronic materials with only short-range order.

We have measured the scattered light from the feather barbs as function of wavelength, polarization, sample orientation, incident angle and observation angle. The cylindrical-shaped feather barbs are mounted at the rotation center of a goniometer, with their axes perpendicular to the rotation axis of the goniometer. Collimated white light is incident on the sample at an angle $\phi$ from the surface normal. Scattered light is collected by a lens and focused to an optical fiber that is connected to a spectrometer. The measured spectra of scattered light are normalized by the spectrum of incident light. The angular resolution of detection is about $5^{\circ}$.
Linear polarizers select the scattered light whose polarization is either parallel (co-polarization) or perpendicular (cross-polarization) to that of incident light.

\begin{figure}[htbp]
\center
\includegraphics[width=3.4in]{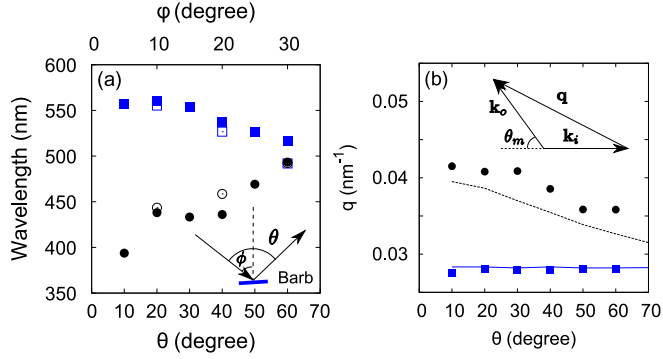}
\caption{(a) Inset: a schematic showing the white light incident on the feather barb at an angle $\phi$ from the surface normal is scattered or reflected to a direction of angle $\theta$ from the incident beam. In the specular reflection measurement, $\theta = 2 \phi$. Main panel:  Wavelengths of the primary (squares) and secondary (circles) peaks in optical scattering spectra (solid symbols) and specular reflection spectra (open symbols) as a function of $\theta$. The top horizontal axis represents the incident angle $\phi$ in the specular reflection measurement.
(b) $q$ values for the primary peak (squares, solid line) and secondary peak (circles, dashed line) as a function of scattering angle $\theta$. The discrete symbols represent experimental data and the lines calculations. Inset: single scattering diagram.}
\label{fig:qm}
\end{figure}

To further confirm the origins of the two peaks, we have measured their angular dependences in a series of experiments. In the first experiment, the direction of incident light and the detector position are fixed. As the sample rotates, the frequencies of the primary peak and secondary peak in the scattering spectra do not change. Their independence of the sample orientation is consistent with the isotropic nature of the nanostructures.
Next we fix the sample orientation and the incident angle of white light, and move the detector to measure the spectra of light scattered into different directions. Both the primary peak and secondary peak shift in wavelength with the detection angle, but in the opposite directions.
We also change the incident angle and repeat the measurement. Analysis of all the data reveals that the frequencies of primary peak and secondary peak do not depend separately on the incident angle or the observation angle, but on their difference, that is, the angle $\theta$ between the directions of illumination and observation [inset of Fig. \ref{fig:qm}(a)]. 
As shown in Fig. \ref{fig:qm}(a), the primary peak shifts to shorter wavelength and the secondary peak to longer wavelength when $\theta$ increases.
We have also measured the reflection spectra of white light for different incident angle $\phi$.  A comparison to the scattering spectra reveals that the reflection peaks at an incident angle $\phi$ have the same frequencies as the scattering peaks at angle $\theta = 2 \phi$ [Fig. \ref{fig:qm}(a)]. Since the angle between the incident beam and the reflected beam is $\theta = 2 \phi$, their spectral coincidence implies the scattering peaks and reflection peaks have the same physical origin. Therefore, specular reflection can be regarded as a special case of light scattering from the quasi-ordered structures.

For single scattering of light from the incident wavevector ${\bf k}_i$ to a wavevector ${\bf k}_o$, ${\bf k}_o - {\bf k}_i = {\bf q}$, where ${\bf q}$ is a spatial vector of the structure. $\theta_m$ is the angle between $-{\bf k}_i$ and ${\bf k}_o$ [inset of Fig. \ref{fig:qm} (b)]. The scattering is elastic,  $|{\bf k}_i| = |{\bf k}_o| \equiv k $.
According to the inset diagram of Fig. \ref{fig:qm} (b), the magnitude of ${\bf q}$ is
\begin{equation}
q= 2 k \cos(\theta_m / 2),
\label{eq:single}
\end{equation}
where $k = 2 \pi n_e / \lambda$, $n_e$ is the effective index of refraction, and $\lambda$ is the vacuum wavelength. We compute the $q$ values for the primary peak and secondary peak in the measured scattering spectra. Light refraction at sample surface is taken into account by converting the experimental angle $\theta$ outside the sample to $\theta_m$ inside the sample \cite{Noh2009}. As shown in Fig. \ref{fig:qm} (b), the $q$ value for the primary peak of light scattering is the same for different $\theta$. It is equal to $q_0$ of the SAXS data [Fig. \ref{fig:pol}(a)] when $n_e$ = 1.25. This value corresponds to a volume fraction of air of $57\%$ in the keratin matrix, which matches the estimate from electron micrographs. Therefore, the primary peak in the optical scattering spectra is produced by single scattering process involving the dominant spatial vectors of the nanostructure.

\begin{figure}[htbp]
\center
\includegraphics[width=3.4in]{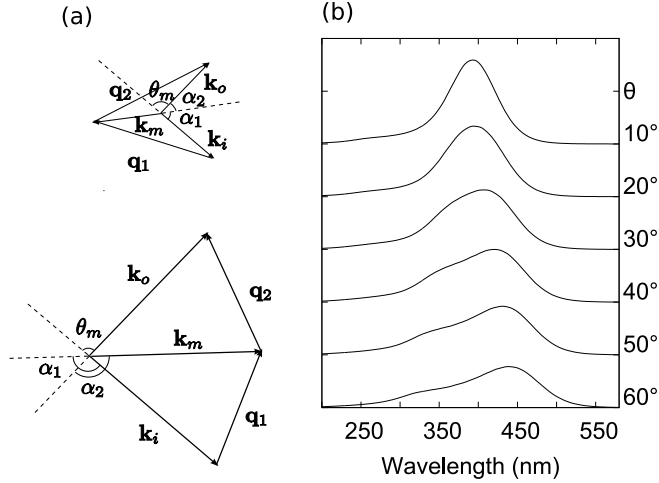}
\caption{(a) Two representative diagrams of double scattering. Light is scattered successively from  ${\bf k}_i$ to ${\bf k}_m$, and then to  ${\bf k}_o$. ${\bf q}_1={\bf k}_m-{\bf k}_i$ and ${\bf q}_2={\bf k}_o-{\bf k}_m$ are spatial vectors of the structure. The angle $\theta_m$ between $-{\bf k}_i$ and ${\bf k}_o$ is the same in the two diagrams.  The single scattering angles $\alpha_1$ and $\alpha_2$ are smaller than 90$^{\circ}$ in the upper diagram and larger than 90$^{\circ}$ in the lower diagram. (b) Calculated double scattering spectra for different scattering angles $\theta$ when the incident angle $\phi$ is fixed at 0$^\circ$. The values of $\theta$ are marked next to the curves.}
\label{fig:sim}
\end{figure}

Unlike the primary peak, the $q$ value for the secondary peak changes with $\theta$. We also compare the optical scattering spectra to X-ray scattering results by plotting the nearly backscattered light intensity ($\theta = 10^{\circ}$) against $q$ in Fig. \ref{fig:pol}(a). While the primary peak of light scattering matches that of SAXS, the secondary peak finds no correspondence in the SAXS data.
This discrepancy is attributed to stronger scattering of light than X-ray by the nanostructures, which results in multiple scattering of light.
The lowest order multiple scattering is double scattering, i.e., the incident light is scattered successively twice before exiting the sample. As shown schematically in Fig. \ref{fig:sim} (a), light is scattered from ${\bf k}_{i}$ to ${\bf k}_m$, and then from ${\bf k}_{m}$ to ${\bf k}_o$. The two successive single scattering events are considered to be uncorrelated.  ${\bf k}_{m}-{\bf k}_i = {\bf q}_1$, and ${\bf k}_{o}-{\bf k}_m = {\bf q}_2$, where ${\bf q}_1$ and ${\bf q}_2$ are spatial vectors of the structure. Light scattering is elastic, $|{\bf k}_{i}|=|{\bf k}_{m}|=|{\bf k}_{o}|= k$.  When the directions of ${\bf k}_{i}$ and ${\bf k}_{o}$ are fixed, ${\bf k}_m$ can be in multiple directions, as illustrated in Figs. \ref{fig:sim}(a). $\alpha_1$ and $\alpha_2$ denote the angles for two successive single scattering processes.  As $\alpha_1$ and $\alpha_2$ increase, $k$ must decrease in order to keep $q_1$ and $q_2$ close to $q_0$.

We calculate the double scattering spectra of light from the Fourier power spectrum $I({\bf q})$ of the structure. $I({\bf q})$, which is measured directly by SAXS, tells the spatial vectors existing in the structure and their strength. Since the structure is isotropic, $I({\bf q}) = I(q)$. For simplicity, we ignore the polarization of light. 
The probability of incident light with ${\bf k}_i$ being scattered to ${\bf k}_m$ is proportional to $I({\bf q}_1 = {\bf k}_m-{\bf k}_i)$, and the probability of successive scattering from ${\bf k}_m$ to ${\bf k}_o$ is proportional to $I({\bf q}_2={\bf k}_o-{\bf k}_m)$. The summation of $I({\bf q}_1) I({\bf q}_2)$ over all possible directions of ${\bf k}_{m}$ gives the probability of incident light being scattered to direction $\theta_m$. In the calculation we sample 1000 directions of ${\bf k}_m$ which are uniformly distributed over $4\pi$ solid angle \cite{Bourke1996}. For comparison with the experimental data, we take into account light refraction at sample surface and sum the scattering spectra over a $5^\circ$ collection angle.

Figure \ref{fig:sim}(b) shows the calculated double scattering spectra when the incident light is normal to the sample surface. As $\theta$ increases, the double scattering peak is broadened and split into two subpeaks. The subpeak at longer wavelength is stronger, and it red-shifts with increasing $\theta$. The broadening of double scattering spectrum can be understood qualitatively by approximating $I(q)$ as a delta function $\delta(q-q_0)$. When $\theta_m$ = 0, $\alpha_1$ and $\alpha_2$ can only be $90^{\circ}$. Only one wavelength is allowed and its value can be obtained by replacing $\theta_m$ in Eq. \ref{eq:single} by $\alpha_1$ or $\alpha_2$. Once $\theta_m > 0$, $\alpha_1$ and $\alpha_2$ can take multiple values. The wavelength is no longer unique. The larger the $\theta_m$, the wider the range of possible values for $\alpha_1$ and $\alpha_2$, and the broader the double scattering spectrum.

Experimentally the efficiency of our detector decreases rapidly when the wavelength of light is shorter than 340 nm. Thus we could not resolve the subpeak at shorter wavelength of the double scattering spectra. In Fig. \ref{fig:qm} (b) we compare the $q$ value for the subpeak at longer wavelength to that of the secondary peak in the measured scattering spectrum. They exhibit similar dependence on $\theta$. Therefore, the secondary peak in the measured light scattering spectra is mostly from the subpeak at longer wavelength of the double scattering spectra in Fig. \ref{fig:sim}(b). As shown in  Fig. \ref{fig:pol}(b), the measured secondary peak in the cross-polarized scattering spectrum is broadened and red-shifted compared to that in the co-polarized one. This is an evidence of triple scattering of light that causes stronger depolarization than double scattering. A qualitative analysis of three successive single scattering events reveals that the triple scattering spectrum is broader than that of double scattering and its center is in between those of single scattering and double scattering.

Note that the secondary peak cannot be generated by high-order diffraction. For instance, the second-order diffraction gives ${\bf k}_o = {\bf k}_i + {\bf q}_1 + {\bf q_2}$. It does not have such a constraint $|{\bf k}_i + {\bf q}_1| = k$, which applies to the double scattering due to $k$ conservation in each of two independent single scattering events. Consequently the second-order diffraction peak is much broader in frequency. For example, in the backward direction ${\bf k}_o = - {\bf k}_i$, with the approximation $I(q) \approx \delta(q-q_0)$, the value of $k$ for the second-order diffraction ranges from $0$ and $q_0$. Hence, the second-order diffraction spectrum spreads from $\lambda = 280 $ nm to infinity, completely different from the measured secondary peak in Fig. \ref{fig:pol}(b).

In summary, we demonstrate that double scattering in a short-range ordered nanostructures produces additional peaks in the light scattering spectrum that contribute significantly to structural colors. This phenomenon has no analog in either periodic structures or random structures. In a perfectly ordered structure, multiple diffraction or high-order diffraction does not create additional peaks,  because a set of primary Bragg peaks would diffract into the same set of primary peaks. Vukusic et al. discovered new colors being generated by double reflection from the concaved multilayers of butterfly wings \cite{vukusica}. The normally incident light, reflected from on one 45$^{\circ}$ inclined surface is directed across the concavity to the opposite orthogonal surface from where it returns parallel to the incident direction. This process strongly depends on the macroscopic geometry, and the direction of incident light.  In contrast, the double scattering phenomenon we report here originates from the intrinsic nanostructure with short-range order. It is insensitive to the surface profile, and exists for any direction of incident light. Therefore, the creation of additional spectral feature by double scattering is a unique property of short-range ordered structures, that has been widely utilized by nature for structural coloration.

This work was supported with seed funding from the Yale NSF-MRSEC (DMR-0520495) and NSF grants to HC (EECS-0823345), ERD (CBET-0547294), and SGJM (DMR), and the W. R. Coe Fund, Yale University (to  ROP). Feather specimens were provided by the Yale Peabody Museum of Natural History and the University of Kansas Natural History Museum and Biodiversity Research Center. SAXS data were collected at beam line 8-ID-I at the Advanced Photon Source at Argonne National Labs with the help of Drs. Alec Sandy and Suresh Narayanan, and supported by the U. S. Department of Energy, Office of Science, Office of Basic Energy Sciences, under Contract No. DE-AC02-06CH11357.

\bibliography{ShortBird}

\end{document}